\documentclass[oneside,english]{revtex4}
\usepackage{amssymb}

\makeatletter

\newcommand{\noun}[1]{\textsc{#1}}

 \newtheorem{thm}{Theorem}[section]
 \newtheorem{defn}[thm]{Definition}

\usepackage{graphicx}

\usepackage{babel}
\makeatother
\begin{document}

\title{{\it CONSTRAINT SATISFACTION BY SURVEY PROPAGATION}} \author{A.
  Braunstein$^{(1,4)}$, M. M\'ezard$^{(2)}$, M. Weigt$^{(3)}$, R.
  Zecchina$^{(4)}$}
\affiliation{ $^{1}$International School for Advanced Studies
  (SISSA),\\ 
  via Beirut 9, 34100 Trieste, Italy \\
  $^{2}$Laboratoire de Physique Th\'eorique et Mod\`eles Statistiques,
  CNRS and Universit\'e Paris Sud, B\^at. 100, 91405 Orsay
  {\sc cedex}, France\\
  $^{3}$ Institute for Theoretical Physics, University of
  G\"ottingen, Tammannstr. 1, 37077 G\"ottingen, Germany \\
  $^{4}$The Abdus Salam International Centre for Theoretical Physics
  (ICTP), \\ Str. Costiera 11, 34100 Trieste, Italy
  \\
}

\begin{abstract}
  Survey Propagation (SP) is an algorithm designed for solving typical
  instances of random constraint satisfiability problems. It has been
  successfully tested on random $3$-satisfiability (3-\noun{sat}) and
  random ${\cal G}(n,\frac{c}{n})$ graph $3$-coloring (3-\noun{col}),
  in the \emph{hard} region of the parameter space, relatively close
  the the SAT/UNSAT phase transition. Here we provide a generic
  formalism which applies to a wide class of discrete Constraint
  Satisfaction Problems.
\end{abstract}
\maketitle

\section{Introduction}
In this paper we suggest a new theoretical framework for the so called
``Survey Propagation'' (SP) equations that are at the root of both the
analysis and the algorithms used in ref.
\cite{MEPAZE,MZ_pre,nosotros,Coloring} to solve the random
3-\noun{Sat} and q-\noun{Coloring} problems. In the more general
context of constraint satisfaction problems we propose a slightly
different way of deriving the equations which we hope can shed some
light on the potentialities of the algorithms and which makes clear
the differences with other well known iterative probabilistic
algorithms. This line of approach, also discussed in \cite{GP} for the
satisfiability problem, is developed here systematically through the
addition of an extra state for the variables which allows to take care
of the clustered structure of the space of solutions.  Within clusters
a variable can be either ``frozen'' to some value -- that is, the
variable takes always the same value for all solutions (satisfying
assignments) within the cluster -- or it may be ``unfrozen'' -- that
is it fluctuates from solution to solution within the cluster.  As we
shall discuss, scope of the SP equations is to properly describe the
cluster to cluster fluctuations by associating to unfrozen variables
an extra state to be added to those belonging to the original
definition of the problem.  The overall algorithmic strategy is
iterative and decomposable in two elementary steps: First, the
marginal probabilities of frozen variables are evaluated by the SP
message-passing procedure; Second -- the so called {\it decimation}
step -- using such information some variables are fixed and the
problem is simplified. While the first step is unavoidable if one is
interested in marginal probabilities, the second step is just dictated
by simplicity and we expect that there could exist other ways of
efficiently using the information provided by the marginals.

Throughout the paper, a detailed comparison with a similar
message-passing procedure, Belief Propagation, which does not make
assumptions about the structure of the solution space will also be
given.

The structure of the paper is as follows.  In Sec. \ref{Generalities},
we provide the general formalism, namely the definitions of Constraint
Satisfaction Problems, Factor Graphs and Cavities, with concrete
reference to the cases of Coloring and Satisfiability.  In
Sec. \ref{BP}, we introduce the {\it warnings} and the {\it local
fields} whose histograms will provide the so called Belief Propagation
equations.  Finally in Sec. \ref{SP}, clusters are introduced and the
SP equations are derived. Explicit equations are given for both
3-\noun{col} and 3-\noun{sat} and the decimation procedure is discussed.

\section{Generalities \label{Generalities}}

\subsection{Constraint satisfaction problems}
 We consider a {\it
  constraint satisfaction problem} (CSP) which is defined on a set of
{\it discrete variables} $\vec x=\left(x_{i}\right){}_{i\in I}$
with $I=\left\{ 1,\dots
  ,n\right\}$. Each variable $x_i$ can be in $q$ possible states (the
generalization to the case where the number of states is $i-$ dependent
is straightforward), so
$\vec x \in
X=\left\{ 1,\dots ,q\right\} {}^{n}$. The vector $\vec x $ is called a 
{\it configuration}.
These variables are subject to a set of {\it constraints}
$\{ C_a \}_{a\in A}$. Each $C_a$ depends on $\vec x$ only
through a  subset $(x_i)_{i\in I(a)}$ of variables. It is
defined as a mapping $C_a:\left\{ 1,\dots ,q\right\} ^{\vert I(a) \vert}\to\{0,1\}
$, where the value $C_a=0$ zero corresponds to a satisfied constraint, and
$C_a=1$ to an unsatisfied constraint. 
It is useful to introduce, for every $i\in I$, the subset $A(i) \subset A$
of indices of all
constraints involving $x_i$. The index sets $I$ and $A$ are chosen
disjoint, so that their elements uniquely determine a single variable
or constraint.

We  define the {\it cost function}
\begin{equation}
  \label{eq:cost}
  C[\vec x]=\sum _{a\in A} C_{a}[\left(x_{i}\right)_{i\in I(a)}]
\end{equation}
which counts the number of unsatisfied constraints. 
Our goal is to
simultaneously satisfy all constraints, i.e. to find a configuration
$\vec s\in X$ with $C[\vec s]=0$. We thus introduce the subset 
$S_C \subset X$ of {\it solutions} to our CSP instance as
\begin{equation}
  \label{eq:sols}
  S_C = \{ \vec s\ |\ \vec s \in X,\ C[\vec s] = 0  \} \ .
\end{equation}

The algorithm  aims at finding one solution $\vec s\in
S_C$. We concentrate {\it a priori} onto instances
$C[\vec x]$ which possess a non-empty solution set $S_C$. 
\subsection{Factor graph}
We  use the {\it factor-graph} \cite{factor_graph} representation
for a CSP:


\begin{defn}
  \label{def:factorgraph}
  For any instance of the CSP problem, we define its {\bf factor
    graph} as a bipartite undirected graph $G=\left(V,E\right)$, having
  two types of nodes:
  \begin{itemize}
  \item variable nodes $i\in I$ and
  \item function nodes $a\in A$.
  \end{itemize}
  Edges connect only different node types; the edge $\left(i,a\right)$  belongs
  to the graph if and only if the constraint  $C_{a}$ 
        involves the variable  $x_{i}$, i.e.  if $a\in
  A\left(i\right)$ or equivalently $i\in I\left(a\right)$. More formally,
  we define $V=A\cup I$ and $E=\left\{ (i,a)\ |\ i\in I,a\in
    A(i)\right\} =\left\{ (i,a)\ |\ a\in A,i\in I(a)\right\} $.
\end{defn}

In figures, we always represent variable nodes by circles, whereas
function nodes are drawn as squares. This notation will help to
distinguish between the different origins of the two node types.

\subsection{Cavities}
 Given a CSP and its factor graph, we will use
the cavity graphs obtained by removing a variable:
\begin{defn}
  \label{def:cavitygraph}
  Given a factor graph $G$ and one variable node $i\in I$, we define the
  {\bf cavity graph} $G^{(i)}$ by deleting from 
  $G$ all function nodes $a\in A(i)$ which are adjacent to $i$, and
  the edges incident to these function nodes.
\end{defn}
The cavity graph $G^{(i)}$  defines a new CSP, where
the cost function is
\begin{equation}
  \label{eq:cavitycost}
  C^{(i)} = C - \sum_{a\in A(i)} C_a = \sum_{b\notin A(i)} C_b \ .
\end{equation}
 Note that in  this new problem
the variable  $x_i$ is isolated, it can take any
value without violating a constraint. 
The solution set $S^{(i)}$ for the cavity problem
 $G^{(i)}$
is larger than the original one, since some constraints have been removed.

\subsection{Two examples: Satisfiability and Coloring}
Although the algorithm can in principle be written for arbitrary CSP,
we shall present two specific examples, satisfiability and coloring.

\begin{figure}
\includegraphics[width=4.cm]{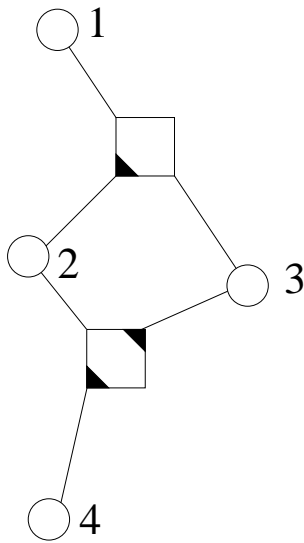}
\caption{The factor graph of a 3-\noun{SAT} problem
corresponding to the simple formula:
  $\left(x_{1}\vee \overline x_{2}\vee x_{3}\right)\wedge
  \left(x_{2}\vee \overline x_{3} \vee \overline x_{4} \right)$.
  Variables are represented as circles, clauses (i.e. function nodes)
  as squares. A triangle-shaped mark indicates that the corresponding
  literal is negated}
\label{cap:sat-fg}
\end{figure}

In the satisfiability problem a constraint $C_a$ is a clause, which
is unsatisfied by only one assignment of the variables $(x_i)_{i \in I(a)}$.
In the random 3-\noun{SAT} problem each clause involves 
three variables ($\vert I(a)\vert=3$), the indices of which are chosen randomly
with uniform distribution in $I$. For a given $a$ and $I(a)$,
 there are eight different types of
constraints $C_{a}$, corresponding to the combinations of possible
negations of literals in one clause, see Fig. \ref{cap:sat-fg}.
In random  3-\noun{SAT} the type of clauses is chosen with uniform distribution 
among these eight types.

\begin{figure}
\includegraphics[width=12.cm]{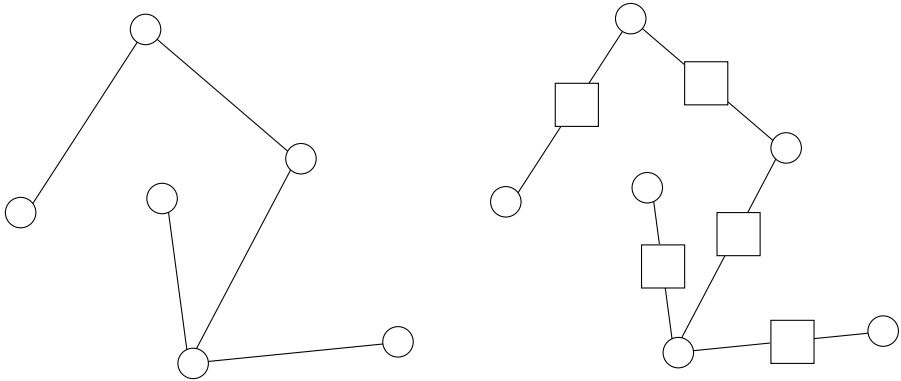}
\caption{The original graph (left) and its factor graph (right)
  corresponding to a coloring problem}
\label{cap:graph-fg}
\end{figure}

In the $q$-coloring problem one is given an original undirected graph.
The problem is to color the vertices, using $q$ colors,
 so that two vertices connected by an edge have different colors.
There is one constraint associated with each edge of the original
graph, and the factor-graph appears as a decoration of the original
graph (see Fig.\ref{cap:graph-fg}), where function nodes have been
added on each original edge. There is only one type of function node.
In the random  $q$-\noun{col} problem, the original graph is a 
 random ${\cal G}(n,\frac cn)$ graph.

We will be particularly interested in the behavior of the algorithm
for large $n$. Note that both $K$-\noun{SAT} and $q$-\noun{col} are
problems where $ \vert A(i) \vert $ have a Poisson limit distribution
with finite mean when $n \to \infty$, i.e.  $ \vert A(i) \vert $ is
typically much smaller than $n$. Moreover, the structure of the factor
graph is locally tree-like. This will guide us in the definition of
the algorithm below, and it is presumably an important ingredient for
the algorithm to work.

\section{Belief propagation \label{BP}}

 \subsection{Warnings and fields}
Given a CSP and a  configuration $\vec x\in X$, we define
the following three quantities associated with $\vec x$,
cf.~\cite{MZ_pre,nosotros}:

\begin{defn}
  \label{def:warning}
  For a given edge $a-i$ of the factor graph, with 
  $i\in I$ and $a\in
  A\left(i\right)$, we define a {\bf warning} as the $q$-component
  vector $\vec u_{a\to i} (\vec x)\in \left\{ 0,1\right\} ^{q}$ with
  components:
  $$
  u_{a\to i}^p (\vec x) = C_{a}\left[ (x_j)_{j\in I(a)}\ |\ x_{i}
    \leftarrow p\right]\ ,\ p=1,...,q,
  $$
  So $ u_{a\to i}^p (\vec x) $ is the value of constraint $C_a$ in 
  the configuration obtained from 
  $\vec x$, by substituting $p$ in the place of  $x_i$.
\end{defn}
\begin{defn}
\label{def:cav_field}  
For a given edge $a-i$ of the factor graph, with 
  $i\in I$ and $a\in
  A\left(i\right)$, we define a
 {\bf cavity field}  as the $q$-component
  vector $\vec h_{i\to a}(\vec x) \in
  \{0,1\}^q$ with components:
  $$
  h_{i\to a}^p(\vec x) =\max _{b\in A(i)\setminus a} u_{b\to i}^p
  (\vec x)
  $$
\end{defn}
\begin{defn}
  \label{def:field}
  For a given node $i\in I$, we define the {\bf local field}
   as the
  $q$-dimensional vector $\vec h_i(\vec x) \in \{0,1\}^q$ with
  components:
  $$
  h_{i}^p(\vec x) =\max _{a\in A(i)} u_{a\to i}^p (\vec x) \ . 
  $$
\end{defn}
 
The warning ${\vec u}_{a\to i}(\vec x)$ is understood as a {\it
  message} sent from constraint $a$ to variable $i$ saying: $x_i$
 cannot be in any of the states $p$  where $u_{a\to i}^p (\vec x)=1$, without
violating constraint $a$. Note that we do not need the value of $x_{i}$
for computing $\vec u_{a\to i}$. In fact, the warning depends
explicitly only on $(x_j)_{j\in I(a)\setminus i}$.

The local field on variable $i$ summarizes all warnings sent to $i$ from the
constraints, i.e. $h_{i}^p(\vec x)\ne 0$ means that,
given the values of all other variables $(x_j), j \ne i$, the variable
$x_i$ should not be assigned the value $p$,
because at least one neighboring constraint would be violated. 

The cavity field  ${\vec h}_{i\to a}(\vec x)$ summarizes all warnings sent
to $i$  from the
constraints different from $a$. 

\subsection{Histograms}
The elementary messages above are defined for an 
arbitrary configuration $\vec x$.
We are  eventually interested in knowing, for 
each variable $i$, the {\it histogram
of local fields} for the configurations which are solutions 
of the CSP: 
\begin{equation}
  \label{eq:Hofh}
  H_i (\vec h) = \frac 1{|S_C|} \sum_{\vec s\in S_C} \delta_{\vec h,
    \vec h_i(\vec s)}\ ,
\end{equation}
 where the ($q$-dimensional)
Kronecker-Delta is simply denoted by $\delta$. This histogram can
also be interpreted as probability distributions $H_{i}(\vec h)=
Prob(\vec h_{i}(\vec s) =\vec h \ | \ \vec s \in S_C)$ of local fields
for randomly chosen solutions.  

Local-field histograms contain useful information
about the set $S_C$ of solutions, which can be exploited
algorithmically in order to recursively construct one solution. If,
e.g., one of the field components is non-zero for all solutions $\vec
s\in S_C$, this particular state is forbidden to this variable. If all
but one components are non-zero, the variable is ``frozen'' to one
specific value in all solutions, i.e. it belongs to the so-called
backbone, and it can be assigned right away.

\begin{figure}[htb]
\includegraphics[width=9.cm]{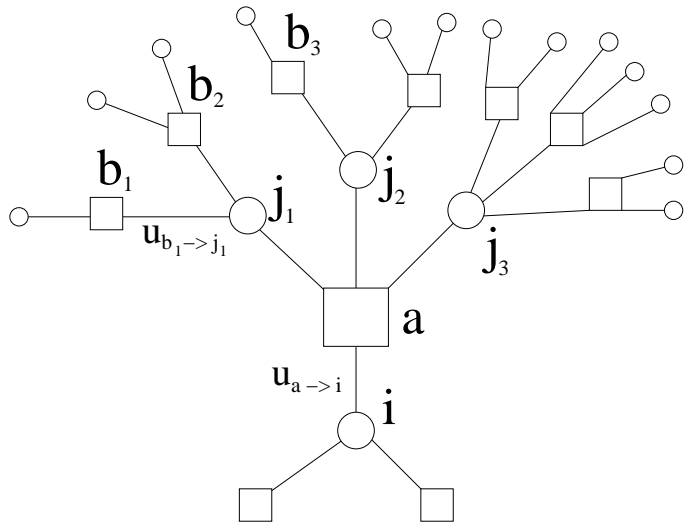}
\caption{Iteration for $\vec u$ warnings.}\label{cap:Iteration}
\end{figure}

Computing $H_i (\vec h)$ is a difficult task, but one may compute it
approximately using a message passing procedure. We first try to find
a recursion relation for the related histograms of the warnings $\vec
u_{a\to i}(\vec s)$ over all solutions $\vec s\in S_C$
\cite{nosotros}. Considering Fig. \ref{cap:Iteration} as an example,
we note that the histogram of $\vec u_{a\to i}(\vec s)$ depends on the
{\it joint} histogram of all the warnings $\vec u_{b\to j}(\vec s)$
sent to all variables $j\in\{j_1,j_2,j_3\}$``above'' function node $a$
(we call them the incoming warnings).  The obvious problem is that
this joint distribution is not known. If the $\vec u_{b\to j}(\vec s)$
were independent variables, we would be able to factorize the joint
histogram into the product of all individual histograms of warnings
$\vec u_{b\to j}(\vec s)$, and then to obtain a recursion. But in
general there is no reason for them to be independent. Moreover, they
cannot even be approximately independent as there are very short paths
joining variables ``above'' variable nodes $j$ (the small unnamed ones
in the figure) between them, variables which in turn define the $\vec
u_{b\to j}(\vec s)$ messages. This is where the cavity graph is
useful.

For each edge $a-i$ of the factor graph, we define the {\it belief}
$B_{a\to i}(\vec u)$ as the histogram of the warning $\vec u_{a\to i}$
over the configurations $s \in S^{(i)}$ which are solutions of the
cavity graph problem:
\begin{eqnarray} 
  B_{a\to i}(\vec u) &=& \frac 1{|S^{(i)}|} \sum_{\vec s\in
    S^{(i)} } \delta_{\vec u, \vec u_{a\to i}(\vec s)}
  \nonumber\\
  &=& Prob\left(\vec u_{a\to i}(\vec s)= \vec u\ |\ \vec s\in
    S^{(i)} \right)\ ,
\label{eq:bare-u}
\end{eqnarray}
where the second line refers again to the probabilistic
interpretation: $B_{a\to i}(\vec u)$ describes the probability of
finding a warning $\vec u$ if a solution of the cavity graph is
randomly selected.

\subsection{Belief propagation equations}\label{bpeq}
Look again at Fig.\ref{cap:Iteration}: If the factor graph $G$ is a
tree, vertices above $j_1,j_2$ and $j_3$ become disconnected if
function nodes $b_i$ are removed, and the various messages $\vec
u_{b\to j}$ are uncorrelated. In this case, we can thus determine the
belief $B_{a\to i}$ as a function of all the incoming beliefs (i.e.
the histograms of the incoming warnings $\{ B_{b\to j} \}$ with $j\in
I\left(a\right)\setminus i$ and $b\in A\left(j\right) \setminus a$),
and so on recursively for the full factor graph.  Standard belief
propagation uses this same recursion also in more general factor
graphs with loops, as a means to compute approximately the local-field
histograms and the beliefs (see e.g.  \cite{YedFreeWei}).

In order to write the corresponding 'belief propagation' equations
explicitly, we use notations similar to those of
Fig.\ref{cap:Iteration}.  Given the edge $a-i$ connecting the function
node $a$ to the variable $i$, we denote by $J$ the set of indices of
the variable nodes ``above'' the function node $a$,
i.e. $J=I\left(a\right)\setminus i$ (in the figure $J=\{ j_1,j_2,j_3\}
$).  For each $j \in J$, we denote by $D_j= A\left(j\right)\setminus
a$ the set of function node ``above'' the variable $j$ (in the figure,
$D_{j_1}=\{ b_1,b_2\} $, and by $\hat D$ the union of these sets:
$\hat D= \bigcup_{j\in } D_j$). The ``incoming messages'', which can
be warning or beliefs, are all the messages propagated on the edges $b
\to j $ with $j \in J$ and, for each such $j$, $b \in D_j$.

Let us first consider a set of incoming warnings $\{\vec u_{b\to
j}\}$.  This warning set may or may not be {}``extensible'' to a
configuration $(s_j)_{j \in J}$ satisfying all constraints $(C_b)_{b
\in \hat D}$.  One can easily go through a bureaucratic procedure to
evaluate all configurations $(s_j)_{j\in J}$ compatible with the
warning set.  First compute the cavity fields \ref{def:field}
component-wise: $h_{j\to a}^p =\max _{b\in D_j} ( u_{b\to j}^p )$.
For each $j \in J$, the allowed values of $s_j$ are those such that
$h_{j\to a}^{s_j} =0$. We denote by $T(\{ \vec {h}_{j\to a} \})
\subset \{1,...,q\}^{\vert J \vert}$ the set of allowed configurations
of the $s_j$ variables:
\begin{equation}
  \label{eq:sols1}
  T(\{\vec h_{j\to a}\}) = \left\{ (s_j)_{j\in I(a)\setminus i } \ |
    \ h_{j\to a}^{s_j} =0, \ \forall j\in I\left(a\right)\setminus i 
  \right\}
\end{equation}
For each $(s_j)$ in $ T(\{\vec h_{j\to a}\})$, one can determine
the output warning $\vec u_{a\to i}$ using definition \ref{def:warning}.

This procedure can be embedded into the probabilistic description of
solutions on the cavity graph. For doing so, we assume that incoming
warnings are {\it independent}.  Following the steps above, one first
calculates from the incoming beliefs the distributions of cavity
fields
\begin{equation}
  \label{eq:cavityHBP}
  H_{j\to a}(\vec h)=\sum _{ \{\vec u_{b\to j} \}_{b\in
      D_j }} \delta_{\vec h , \vec h_{j\to a}} 
    \prod_{b\in D_j} B_{b\to j} (\vec u_{b\to j})\ .
\end{equation}
The new distribution of warnings $\vec u_{a\to i}$ is now given by
an average over cavity fields, 
\begin{equation}
  \label{eq:warningUBP}
  B_{a\to i} (\vec u) 
  = {\cal Z}^{-1} \sum_{\{\vec h_{j\to a}\} _{j\in J }} 
  \left[\sum_{\vec s\in T(\{\vec h_{j\to a}\}) }\delta_{\vec u, \vec
      u_{a\to  i}(\vec s)} \right]
  \prod_{j\in J} B_{j\to a} (\vec
  h_{j\to a}) 
\end{equation}
The prefactor $ {\cal Z}^{-1}$ is a normalization constant. Note that
each cavity-field configuration $\{\vec h_{j\to a}\}$ is contributing
$|T(\{\vec h_{j\to a}\})|$ terms. As a
byproduct, contradictory messages automatically do not contribute
anything to (\ref{eq:warningUBP}).

The BP  equations (\ref{eq:cavityHBP},\ref{eq:warningUBP}) 
are equivalent to the so called sum-product (or
belief network, or Bayesian network) equations
\cite{Gallager,pearl}. One can try to solve them by iteration,
starting form some randomly chosen beliefs, and updating $B_{a \to i}$
sequentially on randomly chosen $a-i$ edges. In 
 some cases, the iteration  converges, independently of the scheme of 
updating, to a unique solution. When the belief
propagation equations converge, one can use the obtained beliefs 
in order to estimate the histogram of local fields,
using:
\begin{equation}
H_j(\vec h) \simeq \sum _{ \{\vec u_{b\to j} \}_{b\in
      A(j) }} \delta_{ \vec h , \vec h_{j\to a}} 
    \prod_{b\in A(j)} B_{b\to j} (\vec u_{b\to j})\ .
\end{equation}
and this histogram can be used for decimation.

\subsection{An example of Belief Propagation: 3-COL}
\begin{figure}
\includegraphics[width=4.cm]{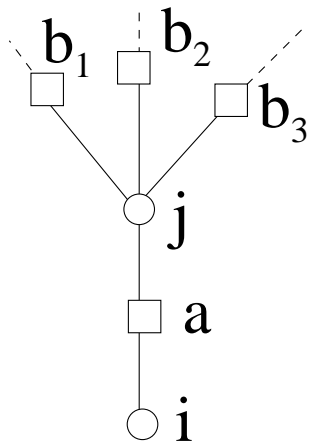}
\caption{An  example of the coloring problem.
This part of the factor graph is the one necessary to compute the messages
(warning and belief)
passed from the function node $a$ to the variable node $i$.}
\label{cap:Coloring-example}
\end{figure}

For the sake of clarity, let us work  out BP on a simple example of the
3-\noun{col} problem ($q=3$), see Fig. \ref{cap:Coloring-example}.
Since function nodes are connected to two variable nodes only
(constraints are edges in the original graph), there is only one
variable node $j$ above function node $a$. For a given configuration
of incoming warnings $\left\{\vec u_{b\to j}\right\}$, we can make a
table of allowed states $s_j$, and for each of them, we can compute
the outgoing warning $\vec u_{a\to i}(s_j)$. Note that possible
warnings are $(1,0,0)$, $(0,1,0)$, $(0,0,1)$, since a function node can
only forbid one color (which is given by the state of the other
variable connected to the function node).
\begin{itemize}
\item Suppose that $\vec u_{b_1\to j}=(1,0,0)$, $\vec u_{b_2\to
    j}=(0,1,0)$, and $\vec u_{b_3\to j}=(0,0,1)$. Then $h_{j\to
    a}=(1,1,1)$, we find a contradictory message. No satisfiable 
  configuration exists for $s_j$. According to the procedure given
  above, this configuration does not contribute. 
\item Suppose that incoming messages are $\vec u_{b_1\to j}= \vec
  u_{b_2\to j} =(1,0,0)$, and $\vec u_{b_3\to j}=(0,1,0)$. Then
  $h_{j\to a}=(1,1,0)$, and the only possible coloring state for $j$
  is $s_j=3$. For this configuration, we thus have only one possible
  outgoing warning, $\vec u_{a\to i}=(0,0,1)$.
\item If $\vec u_{b_1\to j}= \vec u_{b_2\to j} = \vec u_{b_3\to j}
  =(1,0,0)$, then $h_{j\to a}=(1,0,0)$, and there are two possible
  colors for $s_j$, namely states $2$ and $3$. For the first one we
  have $u_{a\to i}=(0,1,0)$, and for the second one $u_{a\to
    i}=(0,0,1)$. Both contribute with equal weight to $B_{a\to
    i}$.
\item All other configurations are simple color permutations of the
  three cases mentioned above, and are handled analogously.
\end{itemize}
From Eqs. (\ref{eq:cavityHBP},\ref{eq:warningUBP}), we can easily
deduce the equation giving the probability distribution $B_{a\to
  i}$ in terms of all distributions $\{ B_{b_l\to j};
l=1,2,3 \}$. Parameterizing $B_{a\to i}$ according to the three
possible messages as
\begin{equation}
  B_{a\to i}(\vec u)=
  \eta_{a\to i}^{1}\delta_{\vec u,(1,0,0)}
  +\eta_{a\to i}^{2}\delta_{\vec u,(0,1,0)}
  +\eta_{a\to i}^{3}\delta_{\vec u,(0,0,1)}\ ,
\label{UBP}
\end{equation}
we find
\begin{equation}
\eta_{a\to i}^{p}=\frac{\prod_{l=1}^3 (1-\eta_{b_l\to j}^p )}
{\sum_{r=1}^q \prod_{l=1}^3 (1-\eta_{b_l\to j}^r)}
\label{etaBP}
\end{equation}
This expression can be easily understood: $\eta_{a\to i}^p$ equals the
probability that color $p$ is forbidden for node $i$, which means that
node $j$ has already taken this color, $s_j=p$. Now, node $j$ can take
color $p$ if and only if it is not forbidden by any incoming warning:
The numerator in Eq. (\ref{etaBP}) simply calculates the probability
that none of the incoming messages forbids color $p$, the denominator
guarantees normalization. Note that configurations in which all
variables above $j$ take the same color $r$, are counted twice, namely
in the expressions for both $p\neq r$. According to the general
discussion given above, this is correct because we have \emph{two} new
configurations for $s_j$, and two corresponding messages $\vec u_{a\to
  i}$ can be sent.

Due to the symmetry between colors, a possible solution would be
$\eta_{a\to i}^p = 1/q$ for all edges $(i,a)\in E$ and all colors $p$.
Note, however, that the main intention is to use these equations in a
recursive coloring algorithm. This means, some variable nodes may
already be assigned a color before, which explicitly breaks the
symmetry. Still, Eq. (\ref{etaBP}) is valid.

\section{Survey propagation \label{SP}}

\subsection{Clustering}

Pitifully, the Belief Propagation dynamics is known not to converge
for the random version of many combinatorial problems (including again
$3$\noun{-sat} and $q$-\noun{col}) in the region of the parameters
near the SAT/UNSAT threshold.  Recently, using tools from statistical
physics, it has been possible to reach some understanding of what
happens in the solution space of these problems around this
threshold~\cite{MEPAZE,MZ_pre,cavity}. Well below the threshold, i.e.
where the number $|A|/n$ of constraints per variable is relatively
small, a generic problem has exponentially many solutions, which tend
to form one giant cluster: For any two solutions, it is possible to
find a connecting path via other solutions that requires short steps
only (each pair of consecutive assignments in the path is close
together in Hamming distance).

Close to the critical threshold, however, the solution space breaks up
into many smaller clusters. Solutions in separate clusters are
generally far apart. In addition, the cost function $C[\vec x]$ has
exponentially many local minima, separated from each other by large
cost ``barriers''. These local cost minima are exponentially more
numerous than the solution clusters, and they act as traps for local
search algorithms.

According to the statistical-physics analysis (which considers the
infinite size limit, $n\to\infty$), there exist exponentially many
widely separated clusters of solutions. Within one such cluster of
solutions, we may identify two types of variables: those which are
frozen in one single state, for all configurations belonging to the
cluster, and those -- unfrozen -- which fluctuate from solution to
solution inside the cluster. Note that also the variables which are
frozen within one solution cluster may change their state when we go
to another cluster, there they may even be unfrozen.  While in general
the above distinction can only provide an approximate description of
clusters, it appears from numerical experiments that in many hard
random CSP, like $K$-\noun{sat} or $q$-\noun{col}, such type of
approximation is already rather accurate.

\subsection{The joker states}
 
Survey propagation turns out \cite{nosotros} to be able to deal with
this clustering phenomenon for large (finite) sizes $n$.  Although the
original derivation uses sophisticated statistical physics ideas, one
can also develop it more directly in algorithmic terms.  The main idea
is that we do not work any more with individual solutions $\vec s\in
S_C$, but with complete clusters of solutions. As already said, some
variables may be frozen inside a cluster, so they retain one single
value $s_j\in \{1,...,q\}$ in our description. Other variables may
take several values n the cluster. For handling these variables, one
can introduce an additional {\it joker state} which we denote by a
``$\star$''. An even finer description, useful for general CSP, uses
varieties of joker states, describing the set of values which are
allowed for the variable, so that $s_j \in {\cal P}$, where $\cal P$
is the ensemble built from all subsets of $\{1,...,q\}$. To each
cluster one should associate exactly one generalized value of each
variable.  One can then generalize the constraint to this enlarged
space and work out the corresponding belief propagation equations. The
resulting equations are the survey propagation equations.

We shall not develop in more details this 'derivation', since it does not
give any rigorous construction, but we will directly
write the equations themselves,  in terms of the original
variables $\vec x \in \{1,...,q\}^n$, and then analyze them.

\subsection{Generalized messages}

We first need to define the generalizations of the warnings, the
cavity-fields and the local field 
used in survey
propagation. In order to lighten the notation
and presentation, we shall drop the 'generalized', and use the same
notation for generalized warning as we used for warnings in the BP
section. The reader should remember that in SP all the messages are
'generalized' messages.

For a given CSP,  we
define the generalized warnings:
\begin{defn}
  \label{def:warningSP}  
  For a given edge $a-i$ of the factor graph, with 
  $i\in I$ and $a\in
  A\left(i\right)$, let $S $ be a given set of possible values
for the variables  $(x_j)_{j \in I(a) \setminus i}$ which are `above' $a$.
We define the {\bf  warning} as the $q$-component
  vector $\vec {\hat u}_{a\to i} (\vec x)\in \left\{ 0,1\right\} ^{q}$ with
  components:
  $$
  \hat u_{a\to i}^p (S) = \min_{(x_j) \in S} 
C_{a}\left[ (x_j)_{j\in I(a)\setminus i } |\ x_{i}
    \leftarrow p\right]\ ,\ p=1,...,q,
$$
\end{defn}
(The generalized warning was called cavity-bias in \cite{MZ_pre}).
Note that the set of possible warnings is enlarged in SP: For the
example of 3-\noun{col} the {\it null message} $(0,0,0)$ is added to
$(0,0,1)$, $(0,1,0)$ and $(1,0,0)$. As we have discussed before, the
non-null messages are sent if the node ``above'' a function node is
assigned a fixed color in the solution cluster.  Correspondingly, the
new message is sent if this vertex is not fixed to a single color,
i.e. if it is in the joker state.

Based on these warnings, we define  local and cavity fields according to
Def. \ref{def:field}, with the argument (a single configuration)
replaced by a set $S$ of configurations: 
\begin{eqnarray}
  \label{eq:clusterfields}
 h_i^p (S) &=& \max _{a\in A(i)} u_{a\to i}^p (S)
 \ ,\nonumber\\ 
 h_{j\to a}^p (S) &=& \max _{b\in A(j)\setminus a }
 u_{b\to j}^p (S)\ . 
\end{eqnarray}

\subsection{Histograms}
Histograms of warnings and fields are now defined as sums over
clusters. The histogram of local fields is given by
\begin{equation} 
  H_i (\vec h) = \frac 1{n_{cl}} \sum_{\alpha=1}^{n_{cl}} 
  \delta_{\vec h, \vec h_i(S_C^{\alpha}) } \ .
\label{eq:cluster-h}
\end{equation}
The histogram of the generalized warning on an edge $a-i$ is now
called the survey, denoted $Q_{a\to i}(\vec u)$.  It is defined in
terms of the clusters of solutions for the cavity graph where $i$ has
been taken away. Calling $S_C^{\alpha,(i)}$ the corresponding
clusters, and $n_{cl}^{(i)}$ their numbers, one defines:
\begin{equation}
  Q_{a\to i}(\vec u) = \frac 1{n_{cl}^{(i)}}
\sum_{\alpha=1}^{n_{cl}^{(i)}}
 \delta_{\vec u, \vec u_{a\to i}(S_C^{\alpha,(i)})} \ .
\end{equation}
 
\subsection{Survey propagation equations}
Based on these definitions, one can easily guess the generalized
recurrence equations for the (approximate) probabilities $Q_{a\to
i}(\vec u)$ that implement the solutions in this enlarged
configuration space. These SP equations lead to a small, yet
fundamental, modification of the BP equations. The basic assumption is
again, that incoming warnings are independent. But contradictory
messages have to be explicitly forbidden. Having
Fig. \ref{cap:Iteration} in mind, and using the same notations as in
sect.\ref{bpeq}, we use the incoming surveys (i.e. the set of surveys:
$\{ Q_{b\to j} \}$ with $j\in I\left(a\right)\setminus i$ and $b\in
A\left(j\right) \setminus a$) to calculate the cavity-field
distributions as in (\ref{eq:cavityHBP}):
\begin{equation}
H_{j\to a} (\vec h) 
= \sum _{\{\vec u_{b\to j} \}_{b\in D_j }} 
\delta_{\vec h, \vec h_{j\to a}}
\prod_{b\in A(j)\setminus a } Q_{b\to j} (\vec u_{b\to j})\ .
\label{eq:cavityH}
\end{equation}
Remember that these fields  may lead to contradictions, if and
only if $\vec h_{j\to a} = (1,1,...,1)$ for at least one $j$. We
therefore introduce the set of all non-contradictory cavity field
configurations,
\begin{equation}
  \label{eq:non-contra}
  {\cal M}_{a\to i} = \left\{ \{\vec h_{j\to a}\}_{j\in I(a)\setminus
      i}\ |\ \forall j: \ \vec h_{j\to a} \in \{0,1\}^q,\ \vec h_{j\to a} \neq
    (1,...,1)\   \right\}\ .
\end{equation}
Then, for an element of ${\cal M}_{a\to i}$ we define again
\begin{equation}
  \label{eq:sols2}
  T\left(\{\vec h_{j\to a}\}\right) = \left\{ (s_j)_{j\in
      I(a)\setminus i } \ |
    \ h_{j\to a}^{s_j} =0, \ \forall j\in I(a)\setminus i 
  \right\}\ ,
\end{equation}
the set of allowed configuration for the variable nodes above
function node $a$. Now the difference to BP enters: All elements of
$T(\{\vec h_{j\to a}\})$ naturally belong to the same cluster, i.e.
they give rise to a {\it single} outgoing warning! 
The new warning is thus 
computed on the set of allowed configurations, it is 
given by $\vec u_{a\to
  i}(T(\{\vec h_{j\to a}\}))$. Its distribution follows immediately, 
\begin{equation} 
  Q_{a\to i}(\vec u)= \tilde{\cal Z}^{-1}
  \sum_{\{\vec h_{j\to a}\} \in {\cal M}_{a\to i}}
  \delta_{\vec u, \vec u_{a\to i}\left(T(\{\vec h_{j\to a}\})\right)}
  \prod_{j\in I(a)\setminus i}H_{j\to a}(\vec h_{j\to a})
  \label{eq:warningUSP}\ .
\end{equation}
The equations (\ref{eq:cavityH},\ref{eq:warningUSP}) are the SP equations.
Note that Eq. (\ref{eq:warningUSP}) produces a dramatic change in the
iteration of the probabilities with respect to the BP Eq.
(\ref{eq:warningUBP}): Every allowed cavity-field configuration
contributes only one term to the sum. Note also that contradictory
messages have to be excluded {\it explicitly} by summing only over
${\cal M}_{a\to i}$. In belief propagation, for each configuration of
input messages one takes the full collection of possible outputs,
thereby introducing a bifurcation mechanism (which may easily become
unstable). On the contrary, in SP the presence of multiple outputs is
collapsed into the null message (which may even not be present in the
belief propagation formalism as it happens for the coloring problem).
A variable which receives a message having at least two zero
components will be ``unfrozen'' in the corresponding cluster.

The SP equations
(\ref{eq:cavityH}-\ref{eq:warningUSP}) provide a closed set of
equation for the surveys. Practically, this recurrence defines a map
\begin{equation}
  \label{eq:Lambda}
  \Lambda :\left\{ ^{old}Q_{a\to i}\right\}
  _{a\in A(i)}^{i=1,\dots ,n}\mapsto \left\{ ^{new}Q_{a\to i}\right\}
  _{a\in A(i)}^{i=1,\dots ,n}
\end{equation}
and we are looking for a fixed point of this map, that will be
obtained numerically by starting with some (random) initial $\left\{
  Q_{a\to i}^0 \right\}$ and applying $\Lambda$ iteratively:
\begin{equation}
  \label{eq:Lambdaiter}
  \left\{ Q^{SP}_{a\to i}\right\}
  =\lim _{m\to \infty }\begin{array}[b]{c} m\textrm{
times}\\ \overbrace{\Lambda \circ \cdots \circ \Lambda }\end{array}
\left\{ Q^{0}_{a\to i}\right\}
\end{equation}
Such a fixed point will be called a ``self-consistent'' set of surveys.

\subsection{An example of Survey Propagation: 3-COL}
For the 3-\noun{col} example, because of the additional null message,
the warning distribution now reads \cite{Coloring_algo}
\begin{equation}
Q_{a\to i}^{SP}(\vec u)=\eta_{a\to i}^{\star}\delta_{\vec u,(0,0,0)}
  +\eta_{a\to i}^{1}\delta_{\vec u,(1,0,0)}
  +\eta_{a\to i}^{2}\delta_{\vec u,(0,1,0)}
  +\eta_{a\to i}^{3}\delta_{\vec u,(0,0,1)} \ ,
\label{USP}
\end{equation}
and the SP equations corresponding to fig.\ref{cap:Coloring-example} are given by
\begin{equation}
  \eta_{a\to i}^{p}=
  \frac{\prod_{l=1}^3 (1-\eta_{b_l\to j}^p) -\sum_{r\neq p}
    \prod_{l=1}^3 (\eta_{b_l\to j}^{\star}+\eta_{b_l\to j}^{r})+
    \prod_{l=1}^3 \eta_{b_l\to j}^{\star}}
  {\sum_{r=1}^q \prod_{l=1}^3 (1-\eta_{b_l\to j}^r)- \sum_{r=1}^q 
    \prod_{l=1}^3 (\eta_{b_l\to j}^{\star}+\eta_{b_l\to j}^r)
    +\prod_{l=1}^3 \eta_{b_l\to j}^{\star}} \ ,
\label{etaSP}
\end{equation}
for $p \in \{1,2,3\}$. Then $\eta_{a\to1}^{\star}$can be computed by
normalization, i.e:
\begin{equation}
  \label{eq:etastarSP}
  \eta_{a\to i}^{\star}=1-\left(\eta_{a\to i}^{1}+\eta_{a\to i}^{2}
    +\eta_{a\to i}^{3}\right)
\end{equation}
The interpretation of this equation is again straightforward, for
simplicity we explain it just for color 1: Now $\eta_{a\to i}^1$ is
given by the probability that $s_j$ is forced to take value $1$, i.e.
by the probability that the cavity field equals $\vec h_{j\to
  a}=(0,1,1)$, conditioned to non-contradictory fields.  The numerator
calculates the unconditioned probability: The first term excludes
$h_{j\to a}^1$ which would forbid color 1 to $s_j$. The second term
takes out fields $(0,h^2,0)$ and $(0,0,h^3)$ where $s_j$ would be
allowed to also take at least one other color. The last term adds
again the probability of field $(0,0,0)$ which was subtracted twice in
the second term.
The denominator realizes the conditioning to non-contradictory fields,
i.e. it gives the probability that $\vec h_{j\to a}\neq (1,1,1)$. The
counting of possible cases follows again the inclusion-exclusion
principle: In the first term, we count fields that have a zero
component in color $r$, summed over $r$. We have to subtract the
double countings due to fields having two zero-components, and we have
to add once the field $(0,0,0)$ which was added three times in the
first term, and subtracted three times in the second one.

Note that the symmetry between colors leads immediately to a solution
$\eta^\star_{a\to i}=1$ for all edges $(i,a)$ of the factor graph,
i.e. only null-messages are sent. This is, however, not the correct
solution in the clustered regime, the color symmetry is not valid at
the level of solution clusters. In fact, the appearance of a
non-trivial solution for the $\eta^p_{a\to i}$ marks the onset of
clustering.

\subsection{The $K$-SAT case}

In the \noun{sat} case, $q=2$, so possible ${\vec u}$ messages are
$\left(0,0\right)$, $\left(1,0\right)$, $\left(0,1\right)$ and
$\left(1,1\right)$.  As any clause can be satisfied by any given
variable (by choosing its value according with the negation of its
corresponding literal) , the $(1,1)$ message will never
show up. Moreover, for a given $a\to i$, which of $(1,0)$ or $(0,1)$
can appear on ${\vec u}_{a\to i}$ will be completely determined by the
sign of the corresponding literal. So we can parameterize distributions
$Q_{a\to i}$ with only one real number $\eta _{a\to i}$ being the
probability of the nontrivial ${\vec u}_{a\to i}$ message
($\left(1,0\right)$ or $\left(0,1\right)$). The probability of $(0,0)$
will simply be $1-\eta _{a\to i}$. The corresponding equations -- which have
been written and implemented in \cite{nosotros,web} -- read in the
case of 3-sat:

\begin{equation}
 \eta_{a \to i}=\prod_{j \in I(a)\setminus i}
\left[
\frac{\Pi^u_{j\to a}}
{\Pi^u_{j\to a}+\Pi^s_{j\to a}+\Pi^0_{j\to a}}
\right]
 \ ,
\label{eta1}
\end{equation}
where:
\begin{eqnarray}
\nonumber
 \Pi^u_{j\to a}&=&
\left[1-\prod_{ b\in A^u_a(j)}\left(1-\eta_{b\to j}\right)\right]
\prod_{ b\in A^s_a (j) }\left(1-\eta_{b\to j}\right) \\
\nonumber
 \Pi^s_{j\to a}&=&
\left[1-\prod_{ b\in A^s_a(j)}\left(1-\eta_{b\to j}\right)\right]
\prod_{ b\in A^u_a (j) }\left(1-\eta_{b\to j}\right)\\
 \Pi^0_{j\to a}&=&
\prod_{ b\in A(j)\setminus a }\left(1-\eta_{b\to j}\right)
\label{eta2}
\end{eqnarray}
$A^u_a(j)$,$A^s_a(j)$ are the two sets in which $A(j)$ is decomposed
($A(j)=A^u_a(j) \cup A^s_a(j)$) where the indexes $s$ (resp:
$u$) refer to the neighbors $b$ for which the literals $(b,j)$ and
$(a,j)$ agree  (resp: disagree). This separation corresponds to the
the distinction of which neighbors  contribute to make variable $j$
satisfy or not-satisfy the clause $a$.

For example, the product $\prod_{ b\in A^s_a(j)}\left(1-\eta_{b\to j}\right)$
gives the probability that no nontrivial message arrives on $j$ from
the function nodes $b\in A^s_a(j)$ (empty products are set to $1$ by
definition).

\subsection{The $q$-COL case}

We have already discussed in detail the 3-\noun{col} problem, a
general number $q$ of colors can be handled analogously. Messages
$\vec u_{a\to i}$ are elements of $\left\{ 0,1\right\} ^{q}$
forbidding the colors which have $1$ in their corresponding
coordinate.  We can immediately see that the possible types of $\vec
u_{a\to i}$ message are $\left(0,\dots ,0,1,0,\dots 0\right)$ (a $1$
on the color taken by the variable in the other end of the link), plus
the additional null message $\left(0,\dots ,0\right)$ (if the
neighboring variable is in the joker state). So we can parameterize
$Q_{a\to i}$ by only $q$ real numbers.

\begin{figure}
\includegraphics[width=5.cm]{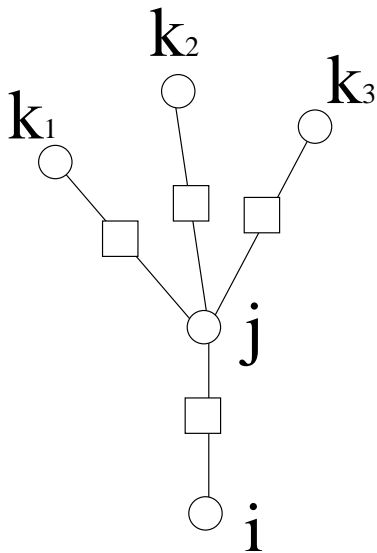}
\caption{Iteration for $q$-\noun{col}}
\label{cap:Iteration-coloring}
\end{figure}

Looking to Figure \ref{cap:Iteration-coloring}, suppose variable $i$
has color $p$ forbidden (i.e. the message $\vec u_{a\to i}$ has a $1$
on component $p$). This implies that on this configuration, variable
$j$ is forced to take color $p$, that is $\vec h_{j\to a}$ is of the
form $\left(1,\dots,1,0,1,\dots ,1\right)$, with a single $0$ in the
$p$-th position ({}``freezing'' type).  For all other $h_{j\to a}$ the
variable $j$ is in the joker state and the output message will be
$\left(0,\dots ,0\right)$.

\subsection{Decimation}

Once the convergence is reached in Eq. (\ref{eq:Lambdaiter}) (we stop
when $\max _{a\in A,i\in I(a)}\left|^{old}Q_{a\to i}-^{new}Q_{a\to
    i}\right|$ becomes small enough), we can use the information
computed so far to find a solution to the original
problem~\cite{MZ_pre,nosotros}. We can easily compute (approximately)
the local-field distributions $\left\{ H_{i}\right\} _{i\in I}$
introduced in Eq. (\ref{eq:cluster-h}) by considering all neighboring
function nodes, and forbidding contradictory messages (remember that
in the cavity graph we have deleted the constraints containing
variable $i$, whereas in $H_i$ we have to restrict the sum to messages
being ``extensible'' to solutions of the complete problem):
\begin{equation}
  H_i (\vec h) = {\cal Z'}^{-1} \sum _{\{\vec u_{a\to i} \}_{a\in A(i)}} 
  \left(1-\delta_{\vec h,(1,...,1)}\right)\ \delta_{\vec h, \vec h_i}
  \prod_{a\in A(i)} Q_{a\to i} (\vec u_{a\to i})\ ,
\label{eq:H}
\end{equation}
with $\vec h_i$ determined according to Eq. (\ref{eq:clusterfields}).

The value $H_i((1,...,1,0,1,...,1))$, with a single zero entry at
component $p$, gives now the probability of a variable $i$ to be
frozen to a certain value $p$. A simple decimation procedure can be
implemented: Select the most frozen variable and fix it to its most
frozen value, then simplify the problem: Certain constraints may
already be satisfied independently of the values of other
participating variables, and can be deleted from the problem instance.
Other constraints may now immediately fix single variables to one
state (unit-clause resolution). Reconverge the warning distributions
on the smaller subproblem.

The decimation algorithm can have three types of behaviors:

\begin{enumerate}
\item The algorithm is able to solve the problem fixing all, or almost
  all variables (some variables may be still unfixed even if the
  problem is already solved).
\item The surveys converge at some stage to the trivial
  solution concentrated on null messages,
  $Q_{a\to i}(u)=\delta_{h,(0,\dots ,0)}$ for all $(i,a)\in
  E$. In this case SP has nothing more to offer. Luckily, these
  problems are generally under-constrained and then easy to solve by
  other means. Note that, for $q$-\noun{col}, the trivial solution
  exists always. In numerical experiments, we found that in case of
  existence of another solution, the latter was the correct one. In
  this case it is therefore reasonable to restart the iteration of the SP equations
  starting from a new random initial condition, even
  if a trivial solution was found once. Only if no non-trivial
  solution can be found after several restarts, the subproblem is passed to a
  different solver.  
\item The SP algorithm does not converge at some stage, even if the
  initial problem was satisfiable.
\end{enumerate}

  On large random instances of $3$-\noun{sat} 
  \cite{nosotros,MZ_pre,MEPAZE,GP} or $q$-\noun{col}
  \cite{Coloring_algo}, in the hard sat region, but not too close to
  the satisfiability threshold, numerical
experiments show that the algorithm behaves as in case 2).

  The generated subproblems turn out to be very simple to solve by
  other conventional heuristics, e.g.  \noun{walksat} \cite{walksat}
  or unmodified belief propagation.

 Case 3) happens in general very close
  to the SAT/UNSAT transition. It is not yet clear if this problem
  appears due to the existence of finite loops in the original problem
  (which make the SP equations to be only approximate), due to the
  simple decimation heuristic which fixes always the most frozen
  variable, or due to some problems which go beyond the validity of
  the SP equation itself.

\section{What's next}

We would like to remark two possible directions of research, among all
those that may follow from the presented algorithm. One is to
formalize rigorously the notions suggested in Section
\ref{SP}, allowing for some well defined definitions of the
clusters, and a corresponding derivation of the SP equations.

Another one, of big computational relevance, is to generalize SP,
which was presented in its purest form, to deal with correlations
between warnings that arise from local problem structures like small
loops in the factor graph, cf. \cite{YedFreeWei} for similar
generalizations of BP. A second possible generalization would include
diverse structures of the space of solutions, e.g. in a sense of
clusters of solution clusters etc.  In the language of statistical
physics, this would include ``more than one step of replica-symmetry
breaking''.

After completing this work, we learned from G. Parisi that he has
reached a similar conclusion on the interpretation of SP in the
colouring problem through the addition of an extra state for the
variables \cite{GP_white}.

{\bf Acknowledgment:} It is a pleasure to thank R. Mulet, A. Pagnani,
and F. Ricci-Tersenghi for numerous discussions. MM and MW acknowledge
the hospitality of the ICTP Trieste where a part of this work was
done. This work has been supported in part through the EC 'STIPCO' network,
grant No HPRN-CT-2002-00319.

\end{document}